\documentclass[10pt, conference]{IEEEtran}
\IEEEoverridecommandlockouts
\usepackage{hyperref}
\usepackage{cite}
\usepackage{amsmath,amssymb,amsfonts}
\usepackage{algorithmic}
\usepackage{graphicx}
\usepackage{textcomp}
\usepackage{xcolor}
\usepackage{textcomp}
\usepackage{stfloats}
\usepackage{url}
\usepackage{verbatim}
\usepackage{booktabs}
\usepackage{amsthm}
\usepackage{subfigure}
\usepackage{makecell}
\usepackage{multirow}
\usepackage{multicol}
\usepackage{enumitem}
\usepackage{subcaption}
\usepackage{color, colortbl}

\definecolor{Gray}{rgb}{0.9, 0.9, 0.9}
\newcommand{\gray}{\cellcolor{Gray}}
\def\model{RedN$^{\text{n}}$D}

\usepackage{float}
\def\BibTeX{{\rm B\kern-.05em{\sc i\kern-.025em b}\kern-.08em
    T\kern-.1667em\lower.7ex\hbox{E}\kern-.125emX}}

\begin{document}

\title{Don't Lose Yourself: Boosting Multimodal Recommendation via Reducing Node-neighbor Discrepancy in Graph Convolutional Network

\thanks{\textbf{$^*$Corresponding Author $^\dagger$Equal Contribution.}}
\thanks{\noindent This work was supported by the National Natural Science Foundation of China (Grant No: 62072390, and 92270123), and the Research Grants Council, Hong Kong SAR, China (Grant No: 15203120, 15226221, 15209922, and 15210023).}}



\author{\IEEEauthorblockN{Zheyu Chen$^{1 \dagger}$, Jinfeng Xu$^{2 \dagger}$, Haibo Hu$^{1 *}$}
\IEEEauthorblockA{The Hong Kong Polytechnic University$^1$, The University of Hong Kong$^2$\\
zheyu.chen@connect.polyu.hk, jinfeng@connect.hku.hk, haibo.hu@polyu.edu.hk}}

\maketitle
\begin{abstract}
The rapid expansion of multimedia contents has led to the emergence of multimodal recommendation systems. It has attracted increasing attention in recommendation systems because its full utilization of data from different modalities alleviates the persistent data sparsity problem. As such, multimodal recommendation models can learn personalized information about nodes in terms of visual and textual. To further alleviate the data sparsity problem, some previous works have introduced graph convolutional networks (GCNs) for multimodal recommendation systems, to enhance the semantic representation of users and items by capturing the potential relationships between them. 
However, adopting GCNs inevitably introduces the over-smoothing problem, which make nodes to be too similar. Unfortunately, incorporating multimodal information will exacerbate this challenge because nodes that are too similar will lose the personalized information learned through multimodal information. To address this problem, we propose a novel model that retains the personalized information of ego nodes during feature aggregation by \textbf{Red}ucing \textbf{N}ode-\textbf{n}eighbor \textbf{D}iscrepancy (\textbf{\model}). Extensive experiments on three public datasets show that \model\space achieves state-of-the-art performance on accuracy and robustness,  with significant improvements over existing GCN-based multimodal frameworks.
\end{abstract}


\section{Introduction}
The rapid expansion of the internet has led to significant information overload, which recommender systems aim to address by predicting user preferences \cite{xu2024improving,xu2024aligngroup}. Although conventional recommender systems have been developed over many years \cite{rendle2012bpr}, the inherent problem of data sparsity still challenges them. To address the data sparsity problem, numerous works have utilized multimodal fusion techniques to enrich the semantic representations of users and items \cite{he2016vbpr,he2017neural,chen2017attentive}. Moreover, graph convolutional networks (GCNs) have been integrated into these models \cite{he2020lightgcn,xu2024fourierkangcf,zhou2023layer} to capture latent relationships between users and items, substantially improving their representational capabilities. Building on these advances, multimodal recommender systems also incorporate GCNs to enrich semantic information \cite{zhou2023tale}. Additionally, recent works have explored the potential of hyper-graph structures \cite{guo2024lgmrec} and diffusion models \cite{jiang2024diffmm} in multimodal recommendation.

However, adopting GCNs inevitably introduces the over-smoothing problem, which lead to nodes to be too similar. And incorporating multimodal information will exacerbate this challenge, because GCNs aggregate information from neighbor nodes, resulting in feature uniformity. Consequently, nodes with numerous similar interactions are treated as almost identical nodes, causing a loss of their personalization. In other words, the over-smoothing problem will cause the model to be unable to fully utilize multimodal information, resulting in suboptimal recommendation performance.  

To address this problem, we propose a novel framework named \textbf{\model}, which retains the personalization of ego nodes by \textbf{Red}ucing \textbf{N}ode-\textbf{n}eighbor \textbf{D}iscrepancy. We refer to the ego nodes with numerous same or similar neighbor nodes as approximate nodes. This method aligns the ego node's neighbors with the ego node, leading to differentiated representations for the same or similar neighbors shared by approximate nodes. It enhances the similarity between the ego node and its neighbors, so that the neighbor representation also has the information of the ego node, increasing the attention of this information in the aggregation process, thereby retaining personalization. This strategy prevents node uniformity, and then alleviates the over-smoothing problem. From the representation perspective, our model reduces excessive clustering of nodes in the feature space, ensuring node representations remain diverse. From the preference perspective, our model allows users to retain their personalization and unique features, reducing the tendency to imitate similar users. Regarding the challenge of this framework: each layer is represented by a representation with the same dimension in GCN, and as layers increase, neighbor nodes have multiple representations, while the ego node has only one. Assigning equal weight to both will lead to neighbors receiving more attention. To address this, we average the representations of neighbor nodes across layers to balance attention between the ego node and its neighbors.

In summary, our key contributions are as follows:
\begin{itemize}[leftmargin=*]
    \item We introduce \model, a GCN-based multimodal framework that mitigates the over-smoothing problem.
    \item We propose a novel alignment task between ego nodes and their neighbors to retain the personalization of ego nodes.
    \item We conduct extensive experiments on three popular datasets, and provide visualization via t-SNE. These results demonstrate the effectiveness and efficiency of \model.
\end{itemize}
\vskip -0.1in
\begin{figure}
    \centering
    \includegraphics[width=1\linewidth]{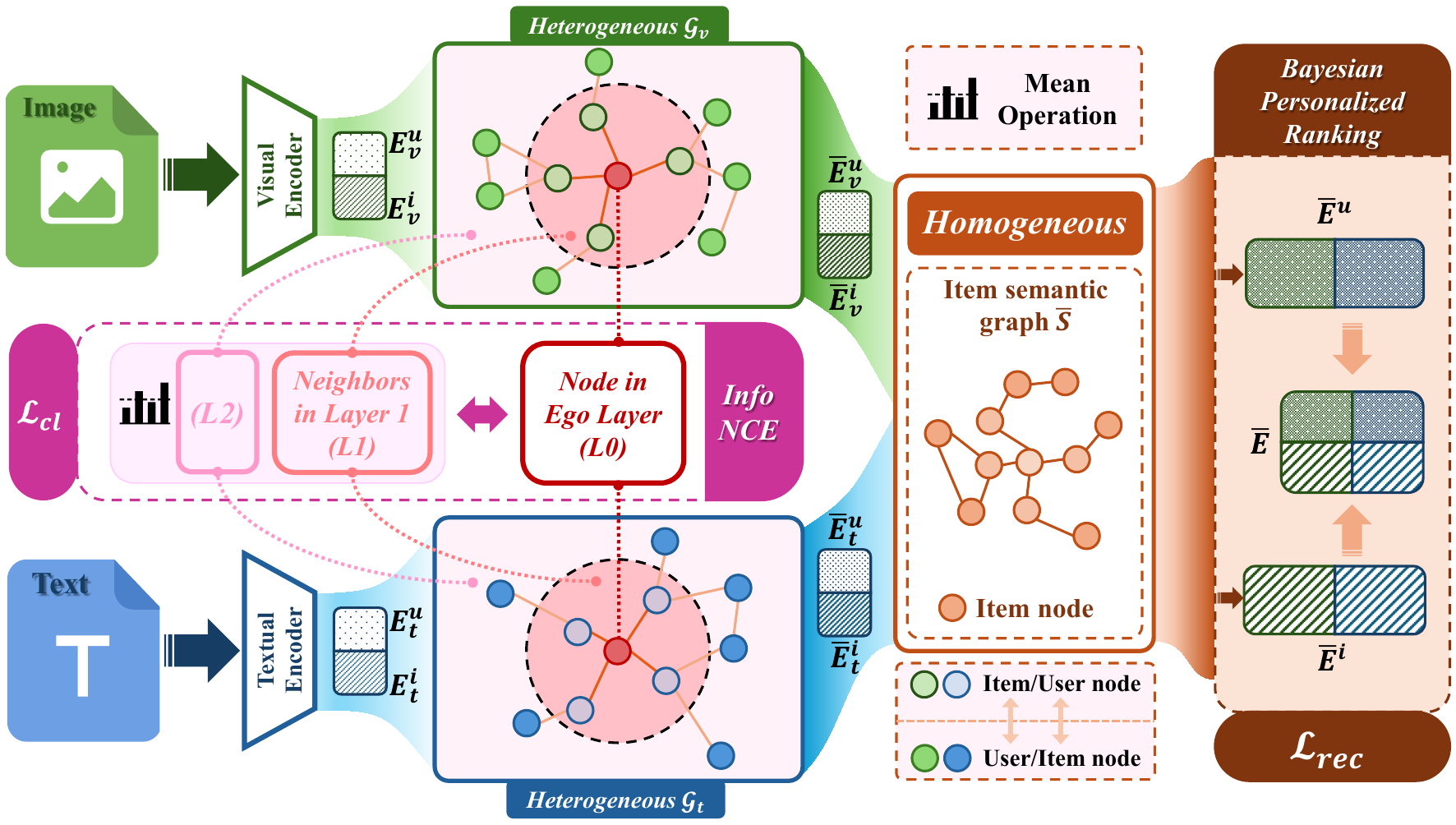}
    \vspace{-2mm}
    \caption{Overall architecture of our \model.}
    \label{overview}
\end{figure}

\section{Methodology}
In this section, we describe each component in \model, and the overall architecture of \model\space is shown in Figure~\ref{overview}.
\vspace{-2mm}
\subsection{Preliminary}
\vspace{-1mm}
Let $\mathcal{U}$ = $\{u\}$ denotes the user set and $\mathcal{I}$ = $\{i\}$ denotes the item set. Then, we denote the features of each modality as $E_{m}$ = $\operatorname{Con}(E^{u}_{m}, E^{i}_{m}) \in \mathbb{R}^{d_m \times (|\mathcal{U}|+|\mathcal{I}|)}$, where $m \in \mathcal{M}$ is the modality. This paper considers visual and textual modalities, denoted as $\mathcal{M}$ = $\{v, t\}$. $d_m$ is the dimension of the features, and $\operatorname{Con}(\cdot)$ denotes concatenation operation.

\vspace{-1mm}
\subsection{Multimodal Information Encoder}
\vspace{-1mm}
Some previous works \cite{zhou2023tale,zhang2021mining,xu2024mentor} find that both the user-item heterogeneous graph and the item-item homogeneous graph can significantly improve the performance of multimodal recommendations. Inspired by them, we propose a multimodal information encoder component to extract the representation of each modality.

\noindent \textbf{Heterogeneous Graph. }
To capture high-order modality-specific features, we construct two \textbf{user-item graphs} $\mathcal{G}$ = $\{\mathcal{G}_{m} \mid m \in \mathcal{M} \}$. Each graph $\mathcal{G}_m$ maintains the same graph structure and only retains the node features associated with each modality. Formally, the message propagation at $l$-th graph convolution layer can be formulated as:
\vskip -0.15in
\begin{equation}
    {E_{m}^u}{(l)}=\sum_{i \in \mathcal{N}_u} \frac{1}{\sqrt{\left|\mathcal{N}_u\right|} \sqrt{\left|\mathcal{N}_i\right|}} {E_{m}^i}{(l-1)}, 
\end{equation}
\vskip -0.12in
\begin{equation}
    E_m^i{(l)}=\sum_{u \in \mathcal{N}_i} \frac{1}{\sqrt{\left|\mathcal{N}_u\right|} \sqrt{\left|\mathcal{N}_i\right|}} E_{m}^u{(l-1)}, 
\end{equation}
\vskip -0.15in
\noindent where ${E_{m}^{u/i}}{(l)}$ represents the user or item representation in modality $m$ at $l$-th graph convolution layer. $\mathcal{N}_{u/i}$ denotes the one-hop neighbors of $u/i$ in $\mathcal{G}$. The final embedding for each modality $m$ is calculated by element-wise summation: 
\vskip -0.1in
\begin{equation}
    \bar{E}_m^{u/i} = \sum_{l=0}^LE_m^{u/i}{(l)}, 
    \label{eq:3}
\end{equation}
\vskip -0.1in
\noindent where $L$ is the number of user-item graph layers. 

\noindent \textbf{Homogeneous Graph. }
We use $k$-NN to establish the \textbf{item-item graph} based on the item features for each modality $m$ to extract significant semantic relations between items. Particularly, we calculate the similarity score $S_m^{i,i^{\prime}}$ between item pair $(i,i^{\prime}) \in \mathcal{I}$ by the cosine similarity $\operatorname{Sim(\cdot)}$ on their modality original features $f^i_m$ and $f^{i^{\prime}}_m$. 
\vskip -0.1in
\begin{equation}
S_m^{i,i^{\prime}}=\operatorname{Sim}(f^i_m, f^{i^{\prime}}_m)=\frac{\left(f^i_m\right)^{\top} f^{i^{\prime}}_m}{\left\|f^i_m\right\|\left\| f^{i^{\prime}}_m\right\|}.
\end{equation}
\vskip -0.1in
\noindent We only retain the top-$k$ neighbors:
\vskip -0.15in
\begin{equation}
\bar{S}_m^{i,i^{\prime}} = \begin{cases} S_m^{i,i^{\prime}} & \text { if } S_m^{i,i^{\prime}} \in \text { top-} k (S_m^{i,h} \mid h \in \mathcal{I}) \\ 0 & \text { otherwise }\end{cases},
\end{equation}
\vskip -0.1in
\noindent where $\bar{S}_m^{i,i^{\prime}}$ represents the edge weight between item $i$ and item $i^{\prime}$ within modality $m$. Thereafter, we further build a unified item-item graph $\bar{S}$ by aggregating all modality-specified graphs $\bar{S}_m$:
\vskip -0.15in
\begin{equation}
\bar{S} = \sum_{m \in \mathcal{M}} \alpha_m \bar{S}_m.
\end{equation}
\vskip -0.1in
\noindent Inspired by \cite{zhou2023tale}, we freeze each item-item graph after initialization to remove the computation costs of the item-item graph during the training phase. Moreover, the $\alpha_m$ is a trainable parameter, which is initialized with equal value for each modality.
\vspace{-1mm}
\subsection{Multimodel Fusion}
\vspace{-1mm}
We calculate the entire user and item representations:
\vskip -0.1in
\begin{equation}
    \bar{E}^{u/i} = \operatorname{Con}(\beta_m \bar{E}^{u/i}_{m} \mid m \in \mathcal{M}),
\end{equation}
\vskip -0.05in
\noindent where the attention weight $\beta_m$ is a trainable parameter, which is initialized with equal value for each modality. Then we enhance $\bar{E}^{i}$ based on $\bar{S}$, the final item-item graph embedding, and fuse visual and textual modalities:
\vskip -0.1in
\begin{equation}
   {\bar{E}}^{i} = {\bar{E}}^{i} + {\bar{S}} \cdot {\bar{E}}^{i}, \quad \bar{E}=\operatorname{Con}(\bar{E}^{u}, \bar{E}^{i}).
\end{equation}

\vspace{-1mm}
\subsection{Node-neighbor Discrepancy Reduction}
\vspace{-1mm}
Equation~\ref{eq:3} shows that as the number of layers in the GCN increases, more layers will be aggregated, making nodes with numerous similar interactions are treated as almost
identical nodes. This aggregating will lead to feature uniformity, causing the over-smoothing problem. To address this problem, our proposed method retain the personalization of ego nodes by reducing the discrepancy between ego nodes and their neighbors.

We also consider that the average strategy will give the neighbor's representation the same attention as the ego node so that the alignment operation is fair and reasonable. Through the analysis above, we obtain $\tilde{e}_m^n$ by averaging the representation of each convolution layer $l$, except the ego layer representation $\hat{e}_m^n$ for modality $m$:
\vskip -0.1in
\begin{equation}
\hat{e}_m^n = e_m^n{(0)}, \quad \tilde{e}_m^n = \frac{1}{L} \sum_{l=1}^{L} e_m^n{(l)},
\end{equation}
\vskip -0.1in
\noindent where $L$ is the total number of convolution layers.

We employ contrastive learning adopting the InfoNCE \cite{oord2018representation} loss function to align the neighbors of the ego node with the ego node. Formally:
\vskip -0.1in
\begin{equation}
    \mathcal{L}_{cl}^m = - \sum_{n \in \mathcal{N}}\log \frac{\exp\Big(\hat{e}^{n\top}_m \tilde{e}^n_m / \tau \Big)} { \sum_{n^\prime \in \mathcal{N}} \exp\Big(\hat{e}^{n\top}_m \tilde{e}^{n^\prime}_m / \tau \Big)},
\end{equation}
\vskip -0.1in
\begin{equation}
    \mathcal{L}_{cl} = \sum_{m \in \mathcal{M}} \mathcal{L}_{cl}^m.
\end{equation}
\vskip -0.1in
\noindent
It enhances the similarity between the neighbors and the ego nodes involved in the aggregation process, thereby strengthening the aggregated information while preserving the unique features of the ego nodes. This strategy effectively prevents node uniformity and mitigates the over-smoothing problem.

\vspace{-1mm}
\subsection{Optimization}
\vspace{-1mm}
We adopt LightGCN \cite{he2020lightgcn} as the backbone model and employ the Bayesian Personalized Ranking (BPR) loss \cite{rendle2012bpr} as the primary optimization objective. The BPR loss is specifically designed to improve the predicted preference distinction between positive and negative items for each triplet $(u, p, n) \in \mathcal{D}$, where $\mathcal{D}$ represents the training dataset. In this context, the positive item $p$ is one with which user $u$ has interacted, while the negative item $n$ is randomly selected from the set of items that user $u$ has not interacted with. Formally:
\vskip -0.2in
\begin{equation} 
    \mathcal{L}_{rec} = \sum_{(u, p, n) \in \mathcal{D}} - \log(\sigma(y_{u,p} - y_{u,n})) + \lambda \cdot \|\mathbf{\Theta}\|^2_2, 
\end{equation}
\vskip -0.1in
\noindent where $\sigma$ represents the sigmoid function, and $\lambda$ controls the strength of $L_2$ regularization, and $\mathbf{\Theta}$ denotes the parameters subject to regularization. The terms $y_{u,p}$ and $y_{u,n}$ correspond to the ratings of user $u$ for the positive item $p$ and the negative item $n$, respectively, computed as $e_u^T \cdot e_p$ and $e_u^T \cdot e_n$. The final loss function is given by:
\vskip -0.2in
\begin{equation}
\mathcal{L} = \mathcal{L}_{rec} + \lambda_c \mathcal{L}_{cl},
\end{equation}
\vskip -0.1in
\noindent where $\lambda_{c}$ is the balancing hyper-parameter.

\section{Experiment}
In this section, we conduct comprehensive experiments to evaluate the performance of our \model\space on three widely used real-world datasets. The following four questions can be well answered through experiment results: \textbf{RQ1}: How does our \model\space model compare to state-of-the-art recommendation models in terms of accuracy? \textbf{RQ2}: What impact do the key components of our \model\space model have on its overall performance? \textbf{RQ3}: How do different hyper-parameters affect the results achieved by our \model\space model? \textbf{RQ4:} Does our \model\space alleviate the over-smoothing problem?

\vspace{-1mm}
\subsection{Datasets and Evaluation Metrics}
\vspace{-1mm}
To assess the performance of our proposed \model\space in the recommendation task, we perform comprehensive experiments on three widely used Amazon datasets \cite{mcauley2015image}: Baby, Sports, and Office. These datasets offer both product descriptions and images. In line with prior works \cite{wei2019mmgcn}, we preprocess the raw data with a 5-core setting for both items and users. Additionally, we utilize pre-extracted 4096-dimensional visual features and obtain 384-dimensional textual features using a pre-trained sentence transformer \cite{zhou2023mmrecsm}.
\begin{table}[h]
    \centering
    \small
\caption{Statistics of datasets.}
\vskip -0.05in
\label{tab:dataset_statistics}
    \begin{tabular}{ccccc}
         \toprule
         \textbf{Datasets}&  \textbf{\#Users}&  \textbf{\#Items}&  \textbf{\#Interactions}& \textbf{Sparsity}\\
         \midrule
         \textbf{Baby} & 19,445 & 7,050 & 160,792 & 99.88\%\\
         \textbf{Sports} & 35,598 & 18,357 & 296,337 & 99.95\%\\
         \textbf{Office} &  4,905 &  2,420 &  53,258 & 99.55\%\\
         \bottomrule
    \end{tabular}
    \vskip -0.1in
\end{table}
For a fair evaluation, we employ two widely recognized metrics: Recall@$K$ (R@$K$) and NDCG@$K$ (N@$K$). We present the average metrics for all users in the test dataset for both $K$ = 10 and $K$ = 20. We adhere to the standard procedure \cite{zhou2023tale} with a random data split of 8:1:1 for training, validation, and testing.

\vspace{-1mm}
\subsection{Baselines and Experimental Settings}
\vspace{-1mm}
To evaluate the effectiveness of our proposed \model, we compare it with state-of-the-art recommendation models, categorized into two groups: conventional recommendation models (\textbf{MF-BPR} \cite{rendle2012bpr}, \textbf{LightGCN} \cite{he2020lightgcn}, \textbf{SimGCL} \cite{yu2022graph}, and \textbf{LayerGCN} \cite{zhou2023layer}) and multimodal recommendation models (\textbf{VBPR} \cite{he2016vbpr}, \textbf{MMGCN} \cite{wei2019mmgcn}, \textbf{DualGNN} \cite{wang2021dualgnn}, \textbf{LATTICE} \cite{zhang2021mining}, \textbf{FREEDOM} \cite{zhou2023tale}, \textbf{SLMRec} \cite{tao2022self}, \textbf{BM3} \cite{zhou2023bootstrap}, \textbf{MMSSL} \cite{wei2023multi}, \textbf{LLMRec} \cite{wei2024llmrec}, \textbf{LGMRec} \cite{guo2024lgmrec}, and \textbf{DiffMM} \cite{jiang2024diffmm}).

We implement \model\space and all baseline models using MMRec \cite{zhou2023mmrecsm}. For the general configuration, we initialize the embeddings using Xavier initialization \cite{glorot2010understanding} with a dimension of 64 and set the learning rate to 1e-4. All models are optimized using the Adam optimizer \cite{kingma2014adam}. To ensure a fair comparison, we conduct a comprehensive grid search for each baseline according to the settings specified in their respective papers. For \model, we perform a grid search over the regularization hyper-parameter $\lambda$ in \{1e-2, 1e-3, 1e-4\}, the balancing hyper-parameter $\lambda_c$ in \{1e-2, 1e-3, 1e-4\}, and the value of $k$ in \{5, 10, 15, 20\} for top-$k$ in constructing the item-item graph. Early stopping is set to 20 epochs to ensure convergence. In line with \cite{zhou2023mmrecsm}, we update the best record based on Recall@20 on the validation dataset.

\begin{table*}[!ht]
\centering
\tabcolsep=0.1cm
\caption{Performance comparison of baselines and \model(our) in terms of Recall@K(R@K) and NDCG@K(N@K). The superscript $^*$ indicates the improvement is statistically significant where the $p$-value is less than 0.01.}
\vskip -0.1in
\label{tab:comparison results}
\begin{tabular}{rcccccccccccc}
\toprule

\multirow{2.5}{*}{\textbf{Model (Source)}}&\multicolumn{4}{c}{\textbf{Baby}} & \multicolumn{4}{c}{\textbf{Sports}} & \multicolumn{4}{c}{\textbf{Office}}\\ \cmidrule(lr){2-5} \cmidrule(lr){6-9} \cmidrule(lr){10-13}&R@10   & {R@20}   & N@10   & {N@20}   & R@10   & {R@20}   & N@10   & N@20    & R@10   & {R@20}   & N@10   &N@20    \\
\midrule
 MF-BPR (UAI'09) & 0.0357& 0.0575& 0.0192& 0.0249
& 0.0432& 0.0653& 0.0241&0.0298
 & 0.0572& 0.0951 & 0.0331&0.0456\\
 LightGCN (SIGIR'20) & 0.0479& 0.0754& 0.0257& 0.0328
& 0.0569& 0.0864& 0.0311&0.0387
 & 0.0791& 0.1189& 0.0459&0.0583\\
 SimGCL (SIGIR'22) & 0.0513 & 0.0804 & 0.0273 & 0.0350 
& 0.0601 & 0.0919 & 0.0327 &0.0414 
 & 0.0799& 0.1239& 0.0470&0.0595\\
 LayerGCN (ICDE'23) & 0.0529& 0.0820& 0.0281& 0.0355& 0.0594& 0.0916& 0.0323&0.0406 & 0.0825& 0.1213& 0.0486&0.0593\\ 
\midrule

{VBPR (AAAI'16)}               &0.0423& 0.0663& 0.0223& 0.0284
& 0.0558& 0.0856& 0.0307& 0.0384
 & 0.0692& 0.1084& 0.0422&0.0531\\
{MMGCN (MM'19)}                 &0.0378& 0.0615& 0.0200& 0.0261
& 0.0370& 0.0605& 0.0193& 0.0254
 & 0.0558& 0.0926& 0.0312&0.0413\\
{DualGNN (TMM'21)}             &0.0448& 0.0716& 0.0240& 0.0309
& 0.0568& 0.0859& 0.0310& 0.0385
 & 0.0887& 0.1350& 0.0505&0.0631\\
{LATTICE (MM'21)}             &0.0547& 0.0850& 0.0292& 0.0370
& 0.0620& 0.0953& 0.0335& 0.0421
 & 0.0969& 0.1421& \underline{0.0562}&\underline{0.0686}\\
{FREEDOM (MM'23)}             &0.0627& \underline{0.0992}& 0.0330& 0.0424
& 0.0717& \underline{0.1089}& 0.0385& \underline{0.0481}
 & \underline{0.0974}& \underline{0.1445}& 0.0549&0.0669\\
{SLMRec (TMM'22)}             &0.0529& 0.0775& 0.0290& 0.0353
& 0.0663& 0.0990& 0.0365&  0.0450
 & 0.0790& 0.1252& 0.0475&0.0599\\ 
{BM3 (WWW'23)}                  &0.0564& 0.0883& 0.0301& 0.0383
& 0.0656& 0.0980& 0.0355& 0.0438
 & 0.0715& 0.1155& 0.0415&0.0533\\
{MMSSL (WWW'23)}               &0.0613& 0.0971& 0.0326& 0.0420
& 0.0673& 0.1013& 0.0380& 0.0474
 & 0.0794& 0.1273& 0.0481&0.0610\\
{LLMRec (WSDM'24)}& 0.0621& 0.0983& 0.0324& 0.0422& 0.0682& 0.1000& 0.0363& 0.0459& 0.0809& 0.1299& 0.0492&0.0621\\
{LGMRec (AAAI'24)}             &\underline{0.0639}& 0.0989& \underline{0.0337}& \underline{0.0430}
& \underline{0.0719}& 0.1068& \underline{0.0387}& 0.0477
 & 0.0959& 0.1402& 0.0514&0.0663\\
{DiffMM (MM'24)}                 &0.0623& 0.0975& 0.0328& 0.0411& 0.0671& 0.1017& 0.0377& 0.0458 & 0.0733& 0.1183& 0.0439&0.0560\\ \midrule
\textbf{\model\space(Our)}&\textbf{0.0663$^*$}& \textbf{0.1039$^*$}& \textbf{0.0361$^*$}&  \textbf{0.0457$^*$}& \textbf{0.0769$^*$}& \textbf{0.1143$^*$}& \textbf{0.0409$^*$}& \textbf{0.0511$^*$} & \textbf{0.1079$^*$}& \textbf{0.1597$^*$}& \textbf{0.0622$^*$}&\textbf{0.0762$^*$}\\
\gray\textit{Improv.\space}&\gray\textit{3.76\%}& \gray\textit{4.63\%}& \gray\textit{7.12\%}&  \gray\textit{6.28\%}& \gray\textit{6.95\%}& \gray\textit{4.96\%}& \gray\textit{5.68\%}& \gray\textit{6.24\%} & \gray\textit{10.78\%}& \gray\textit{10.52\%}& \gray\textit{10.68\%}&\gray\textit{11.08\%} \\\toprule
\end{tabular}
\vskip -0.15in
\end{table*}
\vspace{-1mm}
\subsection{Overall Performance (RQ1)}
\vspace{-1mm}
Detailed experiment results are shown in Table \ref{tab:comparison results}. The optimal results are highlighted in bold, while the suboptimal ones are underlined. Based on these results, we observed that our \model\space outperforms the strongest baselines, achieving 3.76\%(R@10), 4.63\%(R@20) improvement on the Baby dataset, 6.95\%(R@10), 4.96\%(R@20) improvement on the Sports dataset, and 10.78\%(R@10), 9.90\%(R@20) improvement on the Office dataset, which demonstrates the effectiveness of our \model.

\begin{table}[!h]
    \centering
    \caption{Ablation Study.}
    \vskip -0.1in
\label{tab:ablation study}
\begin{tabular}{ccccccc}
\toprule
\multirow{2.5}{*}{\textbf{Variant}}                                 &\multicolumn{2}{c}{\textbf{Baby}}&\multicolumn{2}{c}{\textbf{Sports}} & \multicolumn{2}{c}{\textbf{Office}}\\ \cmidrule(lr){2-3} \cmidrule(lr){4-5} \cmidrule(lr){6-7}
                                  & R@20   & N@20   & R@20   & N@20    & R@20   &N@20    \\
\midrule
{RedN$^1$D}                    &  0.1011&  0.0438& 0.1098& 0.0487 & 0.1576&0.0734\\
{RedN$^2$D}                    & 0.1027& 0.0450& 0.1113& 0.0496 & 0.1581&0.0757\\
{RedN$^3$D}                      &  \textbf{0.1039}&  \textbf{0.0457}& \textbf{0.1143} & \textbf{0.0511} & \textbf{0.1597}&\textbf{0.0762}\\

 {RedN$^4$D}& 0.1031& 0.0452& 0.1137& 0.0506& 0.1593&0.0758\\
 \bottomrule
\end{tabular}
\vskip -0.15in
\end{table}
\vspace{-1mm}
\subsection{Ablation Study (RQ2)}
\vspace{-1mm}
Table~\ref{tab:ablation study} highlights the impact of various excitation strategies. RedN$^1$D means only one neighbor layer of the ego node, and RedN$^2$D, RedN$^3$D, RedN$^4$D means the ego node has two, three, or four neighbor layers. This ablation study demonstrates that the choice of neighbor layer number significantly influences the model's representation capability and performance. Within the experimental range, the three-layer has the best performance.

\vspace{-1mm}
\subsection{Hyper-parameter Analysis (RQ3)}
\begin{figure}[!t]
    \centering
    \subfigure[$k$ in Recall@20] {
        \label{fig:k_R20}
        \includegraphics[width=0.3\linewidth]{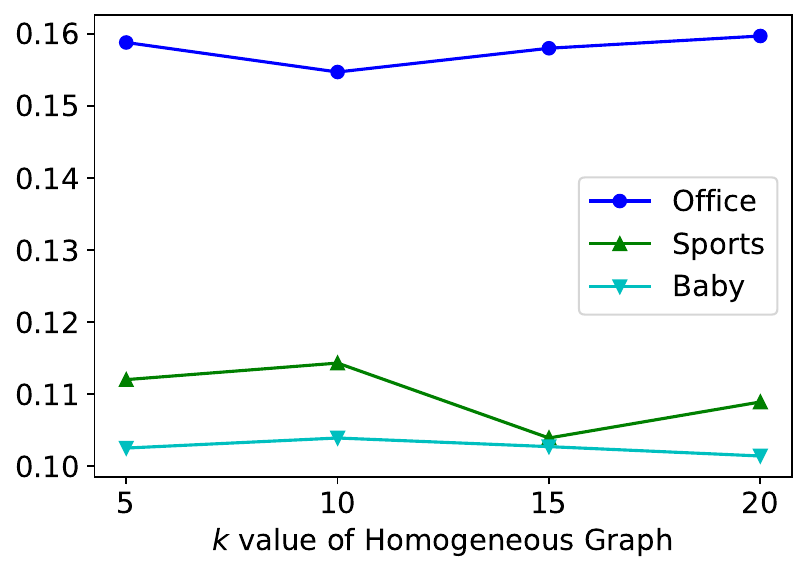}
        } \hspace{-0.15in}
    \subfigure[$\lambda$ in Recall@20] {
        \label{fig:lambda_R20}
        \includegraphics[width=0.31\linewidth]{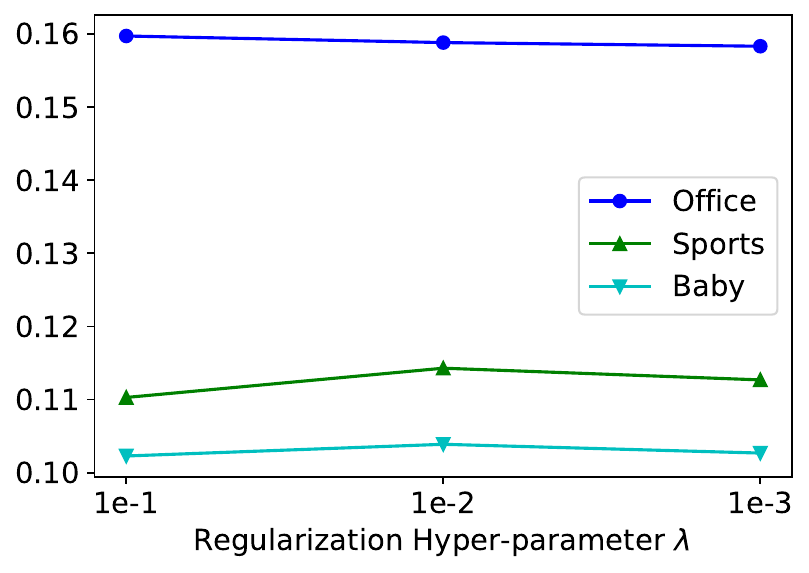}
        } \hspace{-0.15in}
    \subfigure[$\lambda_c$ in Recall@20] {
        \label{fig:lambda_c_R20}
        \includegraphics[width=0.31\linewidth]{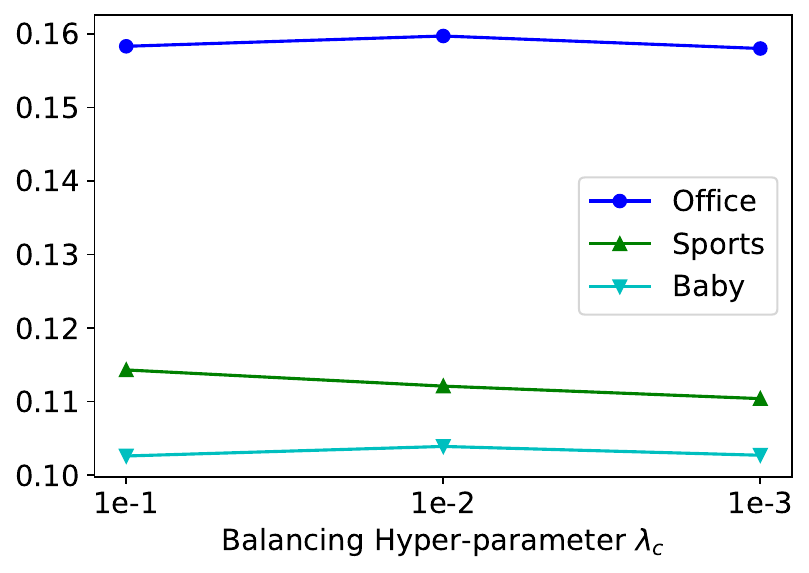}
        } \hspace{-0.15in}
        \vskip -0.1in
    \subfigure[$k$ in NDCG@20] {
        \label{fig:k_N20}
        \includegraphics[width=0.31\linewidth]{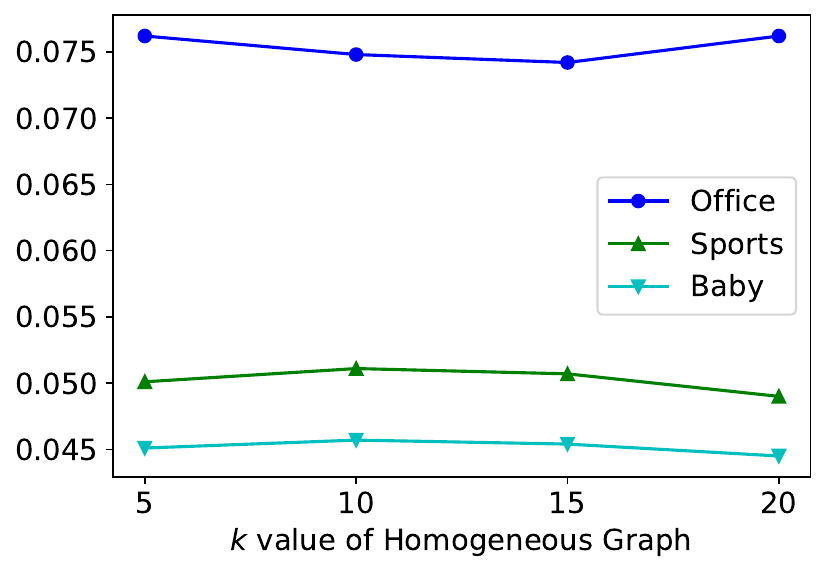}
        } \hspace{-0.15in}
    \subfigure[$\lambda$ in NDCG@20] {
        \label{fig:lambda_N20}
        \includegraphics[width=0.31\linewidth]{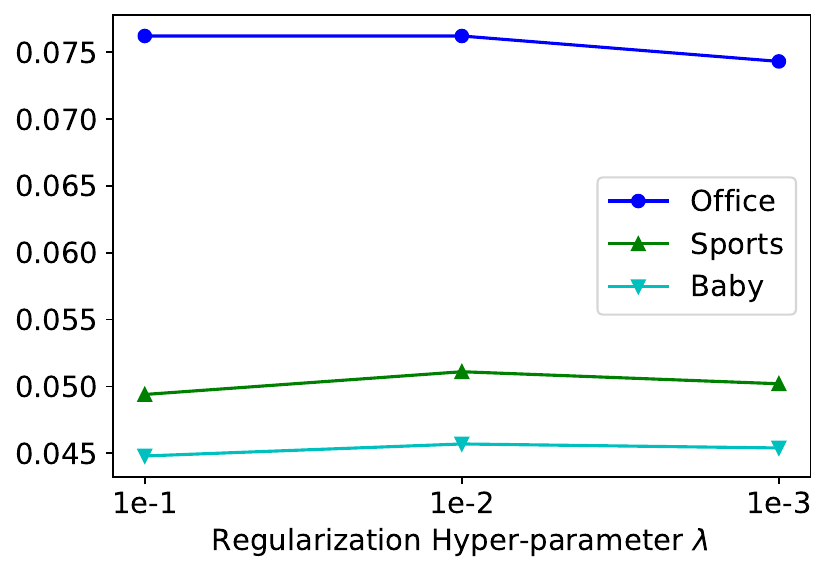}
        } \hspace{-0.15in}
    \subfigure[$\lambda_c$ in NDCG@20] {
        \label{fig:lambda_c_N20}
        \includegraphics[width=0.31\linewidth]{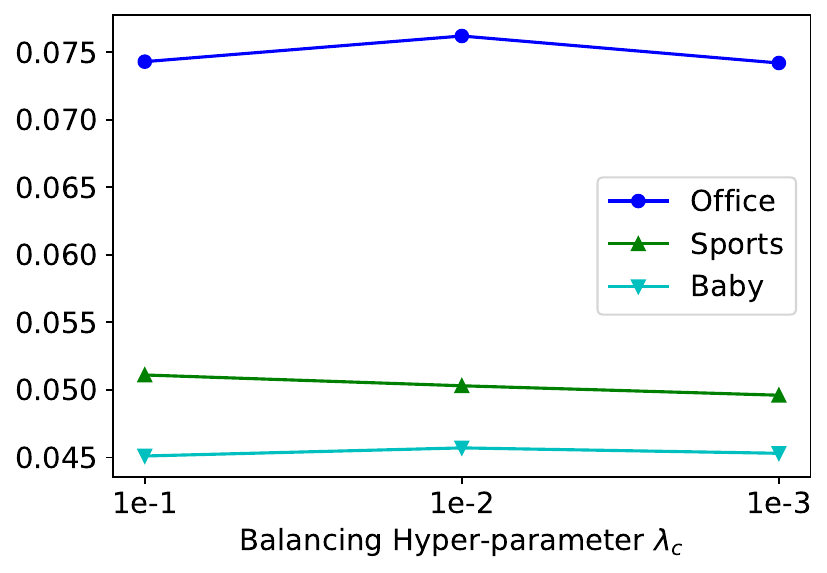}
        }
    \vskip -0.1in
    \caption{Effect of hyper-parameters: $k$, $\lambda$ and $\lambda_c$.}
    \label{fig:hyperparameter}    
    \vskip -0.2in
\end{figure}

\vspace{-1mm}
To examine the sensitivity of \model\space to hyper-parameters, we evaluated its performance on three datasets with different hyper-parameter values. Figure~\ref{fig:k_R20} and Figure~\ref{fig:k_N20} indicate that the optimal $k$-value for constructing the homogeneous graph is 10 for the Baby and Sports datasets, whereas 20 is preferred for the Office dataset. Additionally, Figure~\ref{fig:lambda_R20} and Figure~\ref{fig:lambda_N20} reveal that setting $\lambda$ to 1e-2 yields the best results for the Baby and Sports datasets. In contrast, the Office dataset shows optimal performance at 1e-1, reflecting a different trend compared to the other two datasets. Furthermore, as illustrated in Figure~\ref{fig:lambda_c_R20} and Figure~\ref{fig:lambda_c_N20}, Office and Baby datasets share the same optimal value of 1e-2 for $\lambda_c$, while Sports achieves the best performance at 1e-1.

The Office dataset achieves the best results with a larger $k$ value, indicating that utilizing more neighbors to construct the item-item graph benefits this dataset. This is attributed to the Office dataset's low sparsity, providing more neighbors with similar semantics, thereby improving the effectiveness of the item-item graph. The hyper-parameters $\lambda$ control the strength of $L_2$ regularization, and $\lambda_c$ balancing the attention allocation between self-supervision task and recommendation task. It is worth noting that being flexible in choosing the value of hyper-parameters will allow us to adapt our model to multiple datasets. Although the optimal settings of $\lambda$ and $\lambda_c$ vary, the performance differences are minimal, demonstrating the robustness and stability of \model\space on different data sets.

\begin{figure}[!t]
    \centering
    \subfigure[RedN$^n$D.] {
        \label{fig: align}
        \includegraphics[width=0.4\linewidth]{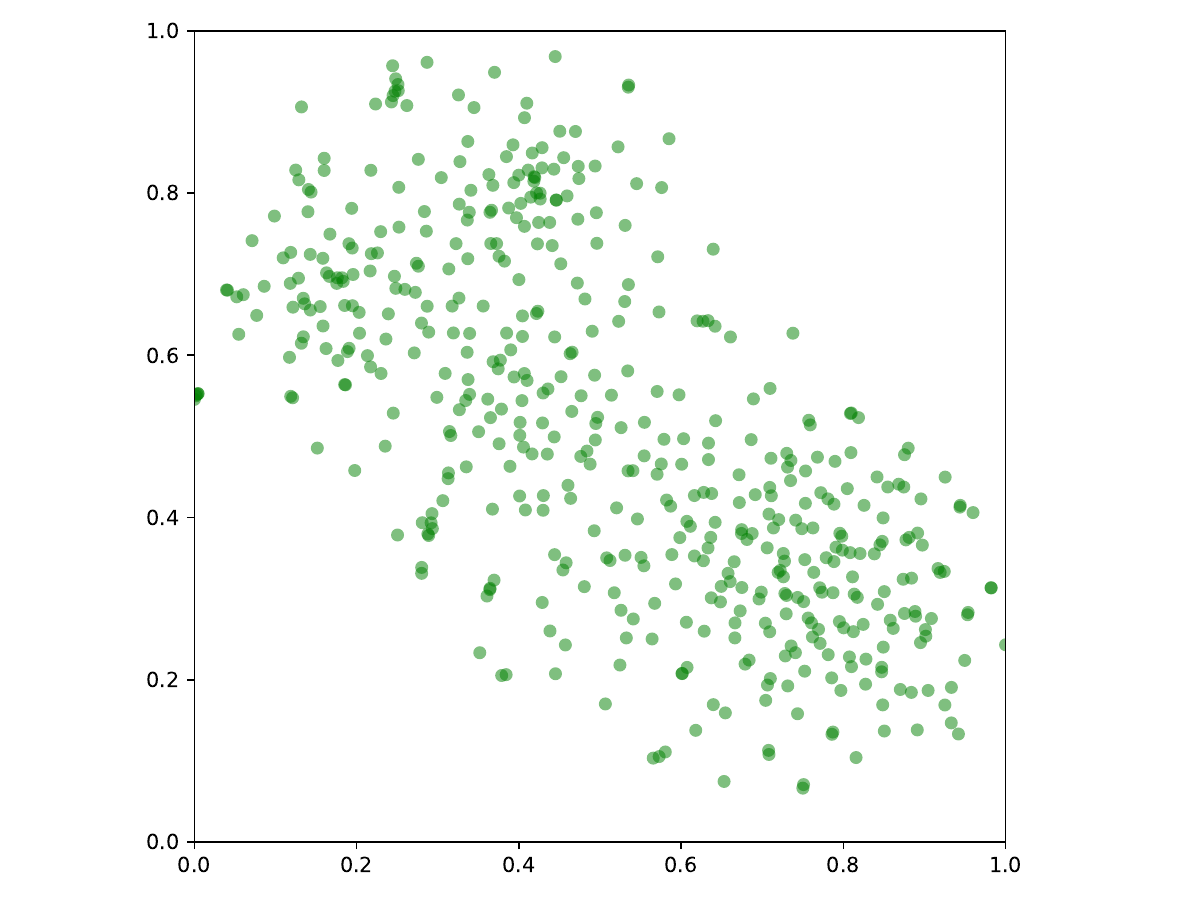}
        }  \hspace{-6.mm}
    \subfigure[RedN$^n$D w/o CL.] {
        \label{fig: origin}
        \includegraphics[width=0.4\linewidth]{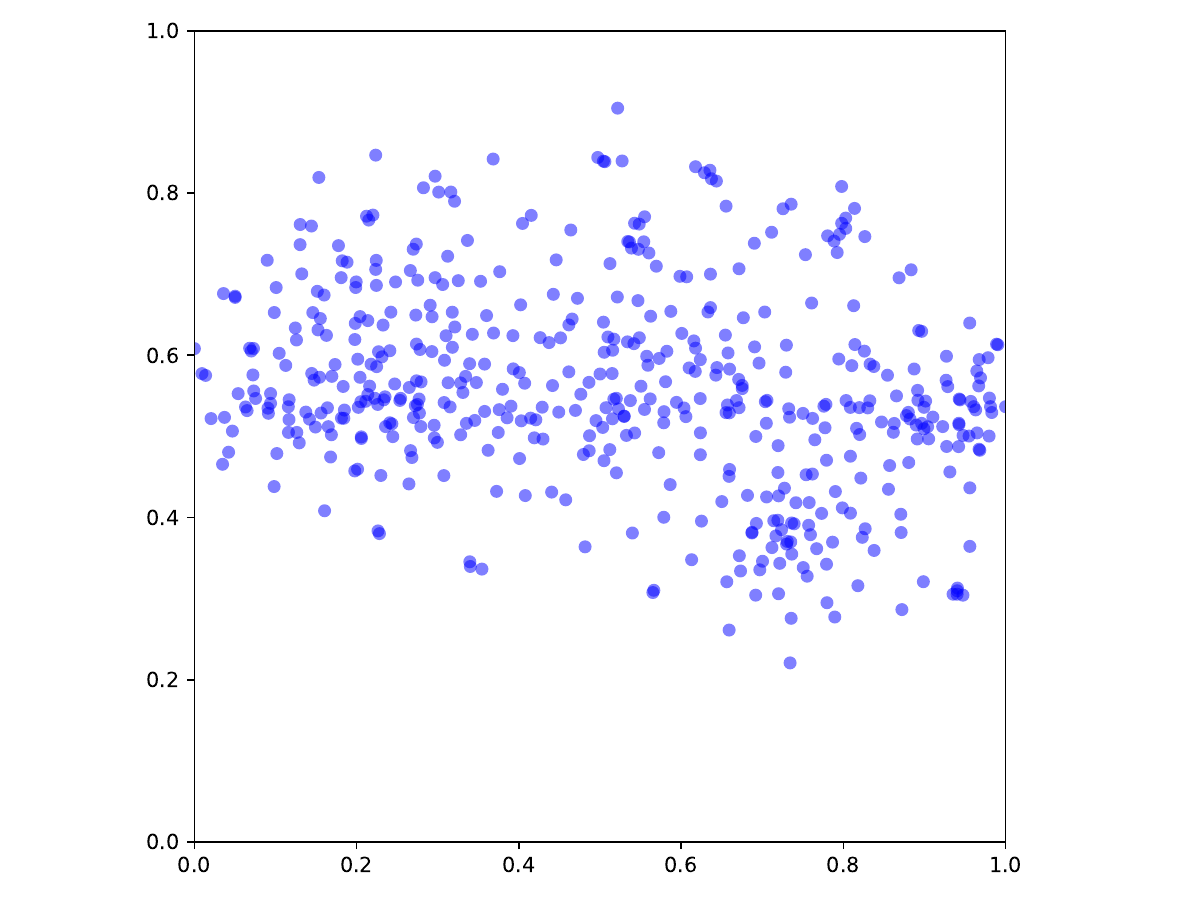}
        }
    \vspace{-2mm}
    \caption{Visualization via t-SNE.}
    \label{fig: t-SNE}
    \vskip -0.2in
\end{figure}

\vspace{-1mm}
\subsection{Visualization (RQ4)}
\vspace{-1mm}
Figure~\ref{fig: t-SNE} visualizes the final 2D embeddings of the optimal model using t-SNE \cite{van2008visualizing}. The green one is the final representation for \model, and the blue one is the final representation for \model\space without CL. The final representation of \model\space is more discrete than that of \model\space without CL, which indicates that the over-smoothing problem is alleviated.

\section{Conclusion}
In this paper, we proposed \model, a novel framework designed to mitigate the over-smoothing problem in GCN-based multimodal recommendation systems. By reducing the discrepancy between ego nodes and their neighbors, \model\space retains the personalization of ego nodes. Comprehensive experiments conducted on three widely used datasets validate the effectiveness of \model, demonstrating significant improvements in recommendation accuracy and robustness compared to existing recommendation frameworks. These results show \model\space mitigates the over-smoothing challenge in GCN-based models, and highlight its potential in advancing the development of multimodal recommendation systems.

\end{document}